\documentstyle[prl,aps,epsf,fancyhdr]{revtex}

\begin{document}

\draft


\title
{Solution of a one-dimensional stochastic model with branching and coagulation reactions\footnote{To appear as a Rapid Communication in Physical Review E}}

\author{Mauro Mobilia and Pierre-Antoine Bares }

\address
{Institute of Theoretical Physics, Swiss Federal Institute of Technology of  Lausanne, CH-1015 Lausanne  EPFL, Switzerland }

\date{\today} 

\maketitle


\begin{abstract}
We solve an one-dimensional stochastic model of interacting particles on a chain. Particles can have branching and coagulation reactions, they can also appear on an empty site and disappear spontaneously.

This model which can be viewed as an epidemic model and/or as a generalization of the {\it voter} model, is treated analytically beyond the {\it conventional} solvable situations. With help of a suitably chosen {\it string function}, which is simply related to the density  and the non-instantaneous two-point correlation functions of the particles, exact expressions of the density and of the non-instantaneous two-point correlation functions, as well as the relaxation spectrum are obtained on a finite and periodic lattice.

\end{abstract}

%
%

%

Because of their important role in the description of classical interacting many-particle non-equilibrium systems, reaction-diffusion (RD) models have been extensively  investigated in the last decade \cite{Privman,Schutzrev}.
In lower dimensions, they provide relevant examples of {\it strongly correlated } systems which cannot be correctly described by mean-field-like approaches. In this sense satisfying comprehension of RD models in lower dimensions would require  {\it exact solutions}, which are scarce, even in one spatial dimension. In some cases, however, certain RD models are known to be solvable. These cases can essentially be classified into four categories:
(i) models for which the equations of motion of correlation functions are closed \cite{Schutz}; (ii) the {\it free-fermion} models \cite{free-fermions} (or systems which can be mapped onto the latter, see \cite{Schutzrev,Henkel}); some other (one-dimensional) RD models can be solved by the (iii) {\it Matrix Ansatz} method \cite{Derrida} and some other by (iv) the {\it interparticle distribution function} (IPDF) method \cite{Doering,Hinrichsen,Peschel}; first introduced for the study of the diffusion-coagulation model (and its variants). It has also to be mentioned that the solution of various one-dimensional RD models have been obtained from the diffusion-coagulation models via {\it similarity transformations}\cite{Schutzrev,Simon}. It has been established that the latter solvable situations correspond to {\it free-fermion} systems \cite{Schutzrev}.

The purpose of this work is to present a generalization of the IPDF method and to apply this technique to solve an one-dimensional stochastic model which is not solvable using {\it conventional} methods. The model under consideration exhibits a {\it massive} spectrum, implying an {\it exponential approach} towards the steady-state. The expressions of the density and non-instantaneous correlation functions are determined.

Consider a periodic  lattice of $L$ sites on which (classical) particles interact. Each site is either empty (denoted by the symbol $\emptyset$) or occupied by a particle at most, say, of species ${\bf A}$ ({\it hard-core interaction}).
When a particle and a vacancy are adjacent to each other, a {\it branching reaction} can take place and the particle ${\bf A}$ can give birth to an offspring (${\bf A}\emptyset\rightarrow {\bf A}{\bf A}$ and $\emptyset {\bf A}\rightarrow {\bf A}{\bf A}$) with rate $\Gamma_{1 0}^{1 1}=\Gamma_{0 1}^{1 1}$; 
another possible reaction is the {\it death} of the particle 
(${\bf A}\emptyset \rightarrow \emptyset\emptyset$ and $\emptyset {\bf A} \rightarrow \emptyset\emptyset$) with rate $\Gamma_{1 0}^{0 0}=\Gamma_{0 1}^{0 0}$. When two particles are adjacent, they can {\it coagulate} (${\bf A}{\bf A}\rightarrow {\bf A}\emptyset$ and ${\bf A}{\bf A}\rightarrow \emptyset {\bf A}$) with rate  $\Gamma_{1 1}^{1 0}= \Gamma_{1 1}^{0 1}$. In addition, when two vacancies are adjacent, a particle can appear ({\it birth} process, $\emptyset \emptyset\rightarrow {\bf A}\emptyset$ and $\emptyset \emptyset\rightarrow \emptyset {\bf A}$) with rate $\Gamma_{0 0}^{1 0}= \Gamma_{0 0}^{0 1}$.
 The system described above can be viewed as an {\it epidemic model} where 
 particles can spontaneously appear/disappear,  have an offspring and coagulate. It can also be viewed as a generalization of the {\it voter} model \cite{Schutzrev}, where the presence/absence of particle is associated to an opinion (yes/no) and each site is associated to an human being. According to the dynamics of the model, each individual changes his opinion at a rate proportional to the opinion of his neighbours.

A particle (vacancy) at each of the $L-$lattice sites corresponding to spin down (up), the master equation of the model can be rewritten as an imaginary-time Schr\"odinger equation for a {\it quantum spin-chain} problem: $\frac{\partial}{\partial t}|P(t)\rangle = -H |P(t)\rangle$, where $|P(t)\rangle=\sum_{\{n\}}P(\{n\},t)|\{n\} \rangle$ describes the state of the system at time $t$ (the sum runs over the $2^{L}$ configurations) and $H$ is the {\it stochastic Hamiltonian} (non hermitian) expressed in a spin-$\frac{1}{2}$ representation as
 $H=\sum_{j=1}^{L}H_{j,j+1}$, with
\begin{eqnarray}
\label{eq.0.1}
-H_{j,j+1}&=&\Gamma_{1 0}^{0 0}\left\{(1-n_{j+1})(\sigma_{j}^{+}-n_{j})+(1-n_{j})(\sigma_{j+1}^{+}-n_{j+1})\right\}\nonumber\\
&+&\Gamma_{0 0}^{1 0}\left\{(1-n_{j+1})(\sigma_{j}^{-}+n_{j}-1)+(1-n_{j})(\sigma_{j+1}^{-}+n_{j+1}-1)\right\} \nonumber\\
&+&\Gamma_{1 1}^{1 0}\left\{n_{j}(\sigma_{j+1}^{+}-n_{j+1})+n_{j+1}(\sigma_{j}^{+}-n_{j})\right\}\nonumber\\
&+&\Gamma_{1 0}^{1 1}\left\{n_{j+1}(\sigma_{j}^{-}+n_{j}-1)+n_{j}(\sigma_{j+1}^{-}+n_{j+1}-1)\right\},
\end{eqnarray}
where the $\sigma^{\pm}$ are the usual Pauli matrices and $n_{j}\equiv\frac{1}{2}(1-\sigma_{j}^{z})$ is the density operator at site $j$. We also define the ``left vacuum'' $\langle \widetilde{\chi}|\equiv \sum_{\{n\}}\langle\{n\}|$. The probability conservation yields $\langle \widetilde{\chi}|H=0$.

For the model under consideration, the equation of evolution of the density, from an initial state $|P(0)\rangle$, is therefore:
\begin{eqnarray}
\label{eq.0.2}
\frac{d}{dt}\langle n_{j}\rangle=-\langle\widetilde{\chi}| n_{j}(H_{j-1,j}+H_{j,j+1})e^{-Ht}|P(0)\rangle= 2A+B\left(\langle n_{j+1}\rangle+\langle n_{j-1}\rangle\right)-2C\langle n_{j}\rangle + D\left(\langle n_{j}n_{j+1}\rangle+\langle n_{j} n_{j-1}\rangle\right),
\end{eqnarray}
where $A\equiv\Gamma_{0 0}^{1 0},
 B\equiv\Gamma_{1 0}^{1 1}-\Gamma_{0 0}^{1 0}, C\equiv\Gamma_{1 0}^{0 0}+\Gamma_{0 0}^{1 0}$ and $D \equiv \Gamma_{1 0}^{0 0}+\Gamma_{0 0}^{1 0}-(\Gamma_{1 0}^{1 1}+\Gamma_{1 1}^{1 0})$. 
 When $D=0$ and $B\neq C$, for a translationally-invariant system with initial density of particles $\langle n_{j}(0)\rangle=\rho(0)$, the solution of (\ref{eq.0.2}) simply reads $\langle n_{j}(t)\rangle=\frac{A}{C-B}+\left(\rho(0)-\frac{A}{C-B} \right)e^{-2(C-B)t}$. However, when $D\neq 0$, it is clear from (\ref{eq.0.2}) that the equation of motion of the correlation functions of the model give rise to an open hierarchy \cite{Schutz,Schutzrev}, which is not in general solvable. In addition, the Hamiltonian (\ref{eq.0.1}) is not quadratic and cannot, in general (excepted when $\Gamma_{1 0}^{1 1}=\Gamma_{1 0}^{0 0}$ and $\Gamma_{1 1}^{1 0}=\Gamma_{0 0}^{1 0}$, see \cite{Schutzrev} for a complete classification of {\it free-fermion} systems), be casted into a free-fermion form. Furthermore, this model cannot be solved by the {\it traditional} IPDF method which is not applicable  \cite{Doering,Hinrichsen,Peschel} in the presence of the processes ${\bf A} \emptyset \rightarrow \emptyset \emptyset\;;\; \emptyset {\bf A} \rightarrow \emptyset \emptyset $ and in the absence of processes  ${\bf A} \emptyset\rightarrow \emptyset {\bf A}\;;\; \emptyset {\bf A} \rightarrow {\bf A} \emptyset $ (the latter should occur with the same rate as the coagulation rates \cite{Doering,Hinrichsen,Peschel}).

 To our knowledge,  no {\it exact results} have been obtained for the model under investigation beyond the $D=0$ and {\it free-fermion} cases. In order to obtain the exact expression of the density beyond the latter {\it conventional cases}
, we generalize the IPDF method introducing the following {\it string function} ($x+L\geq y\geq x\geq 1$):
\begin{eqnarray}
\label{eq.0.3}
S_{x,y}(t)\equiv\langle(a-bn_{x})(a-bn_{x+1})\dots(a-bn_{y-1})\rangle(t),
\end{eqnarray}
where $a$ and $b$ are non-vanishing numbers.
 When $a=b=1$, $S_{x,y}(t)$ is the {\it empty interval functions} used in  the {\it traditional} IPDF method \cite{Doering,Hinrichsen,Peschel}. 
The idea to solve the model under consideration here (with certain restrictions for the reaction-rates) is to {\it choose} suitable $a$ and $b$ in order to have a closed equation of evolution for $S_{x,y}(t)$. This is achieved by imposing the following ratio between $a$ and $b$
\begin{eqnarray}
\label{eq.0.4}
\frac{b}{a}=1+\frac{\Gamma_{1 1}^{1 0}}{\Gamma_{0 0}^{1 0}}>1
\end{eqnarray}
and for the following reaction rates:
\begin{eqnarray}
\label{eq.0.5}
\Gamma_{1 1}^{1 0}=\Gamma_{1 1}^{0 1}>0;\;\; 2\Gamma_{0 0}^{1 0}= 2\Gamma_{0 0}^{0 1}\geq
\Gamma_{1 0}^{1 1}=\Gamma_{0 1}^{1 1}\geq 0;\;\; \text{and}\;\;
\Gamma_{1 0}^{0 0}=\Gamma_{0 1}^{0 0}=\frac{\Gamma_{1 1}^{1 0}(2\Gamma_{0 0}^{1 0}-\Gamma_{1 0}^{1 1})}{\Gamma_{0 0}^{1 0}}\geq 0
\end{eqnarray}
According to (\ref{eq.0.4}) and with the rates (\ref{eq.0.5}), for the model under consideration on a periodic lattice of $L$ sites, we have ($1\leq x\leq y\leq x+L$):
\begin{eqnarray}
\label{eq.0.6}
\left\{
\begin{array}{l l l}
\frac{d}{dt}S_{x,y}(t)=\frac{\alpha}{2}\left(S_{x+1,y}(t)+S_{x,y-1}(t)\right)+
\frac{\beta}{2}\left(S_{x-1,y}(t)+S_{x,y+1}(t)\right)-\gamma S_{x,y}(t)-(y-x)\delta S_{x,y}(t)
;\;\; (1\leq x<y<x+L)\\
\frac{d}{dt}S_{x,x+L}(t)=-L\delta S_{x,x+L}(t)\\
S_{x,x}(t)=1,
\end{array}
\right.
\end{eqnarray}
where $\alpha\equiv 2(aC-bA), \beta\equiv-2D/b, \gamma\equiv 2(B+C)-\delta$ and $\delta\equiv \frac{2b}{a}A> 0$. The prescription $S_{x,x}(t)=1$ is obtained requiring that
 $S_{x,x+1}(t)=a-b\langle n_{x}(t)\rangle$ and using (\ref{eq.0.2}).

 The subcase $\Gamma_{1 0}^{1 1}=\Gamma_{0 0}^{1 0}$ implies $\alpha=\beta=B=D=0$ and we recover ($C\neq 0$) $\langle n_{x}(t)\rangle =\frac{a-S_{x,x+1}(t)}{b}=\frac{A}{C}+\left(\langle n_{x}(0)\rangle-\frac{A}{C}\right)e^{-2Ct}$.

 Hereafter we focus on the more general situation where (\ref{eq.0.5}) are fulfilled with   $\Gamma_{1 0}^{1 1}\neq \Gamma_{0 0}^{1 0}$, and thus $\alpha\neq 0, \beta\neq 0$ . 

 It is useful to consider the {\it auxiliary} function  ${\cal R}_{x,y}(t)\equiv \mu^{x-y}S_{x,y}
(t)$, where we introduce the complex numbers $\mu\equiv -i sgn(\alpha) |\frac{\alpha}{\beta}|^{1/2}$ and 
$q\equiv i |\alpha \beta|^{1/2}\neq 0$. Notice that, because of (\ref{eq.0.5}), $0<\frac{|q|}{\delta}<\frac{1}{2}$.  With help of (\ref{eq.0.5}), we obtain the equation of motion of ${\cal R}_{x,y}(t)$:
\begin{eqnarray}
\label{eq.0.12}
\left\{
\begin{array}{l l}
\frac{d}{dt}{\cal R}_{x,y}(t)=\frac{q}{2}\sum_{e=\pm1}\left\{ {\cal R}_{x+e,y}(t)+ {\cal R}_{x,y+e}(t) \right\}-\gamma {\cal R}_{x,y}(t)-(y-x)\delta {\cal R}_{x,y}(t);\;\; (1\leq x<y<x+L)\\
\frac{d}{dt}{\cal R}_{x,x+L}(t)=- L\delta {\cal R}_{x,x+L}(t)\\
{\cal R}_{x,x}(t)=1,
\end{array}
\right.
\end{eqnarray}
The stationary solution of (\ref{eq.0.12}) is obtained with the Ansatz: 
${\cal R}_{x,y}(\infty)=\widetilde{A}_{L}J_{y-x+\omega}(2q/\delta)+
\widetilde{B}_{L}Y_{y-x+\omega}(2q/\delta)$,where $J_{\nu}(z)$
 and $Y_{\nu}(z)$ are the usual Bessel functions of first and second kind,
 respectively, $\widetilde{A}_{L}$ and $\widetilde{B}_{L}$ are constants to 
be determined. Inserting the expression of
  ${\cal R}_{x,y}(\infty)$ into (\ref{eq.0.12}), we obtain
 $\omega=\gamma/\delta$. Taking into account the boundary conditions 
${\cal R}_{x,x}(t)={\cal R}_{x,x}(\infty)=1$ and ${\cal R}_{x,x+L}(\infty)=0$,
 we get
\begin{eqnarray}
\label{eq.0.12.0}
\widetilde{A}_{L}&=&-\frac{Y_{L+\gamma/\delta}(2q/\delta)}{J_{L+\gamma/\delta}(2q/\delta)
Y_{\gamma/\delta}(2q/\delta)-Y_{L+\gamma/\delta}(2q/\delta)
J_{\gamma/\delta}(2q/\delta) } \\
\widetilde{B}_{L}&=&\frac{J_{L+\gamma/\delta}(2q/\delta)}{J_{L+\gamma/\delta}(2q/\delta)
Y_{\gamma/\delta}(2q/\delta)-Y_{L+\gamma/\delta}(2q/\delta)
J_{\gamma/\delta}(2q/\delta) },
\end{eqnarray}
which provides the stationary expression for the string function:
\begin{eqnarray}
\label{eq.0.12.1}
S_{x,y}(\infty)=\mu^{y-x}\left(\widetilde{A}_{L}J_{y-x+\gamma/\delta}(2q/\delta)
+\widetilde{B}_{L}Y_{y-x+\gamma/\delta}(2q/\delta)
\right)
\end{eqnarray}
According to the definition of the string function, the density of particles at site $x$ is given by $\langle n_{x}(t)\rangle=\frac{a-S_{x,x+1}(t)}{b}$ and therefore the explicit stationary density of particles  reads
\begin{eqnarray}
\label{eq.0.13}
\langle n_{x}(\infty)\rangle=\frac{a-S_{x,x+1}(\infty)}{b}=\frac{1}{b}\left[a-\mu\left(\widetilde{A}_{L}J_{1+\gamma/\delta}(2q/\delta)+ 
\widetilde{B}_{L}Y_{1+\gamma/\delta}(2q/\delta)
\right)\right].
\end{eqnarray}
In order to solve the dynamical part of (\ref{eq.0.12}), we seek a solution of the form ${\cal R}_{x,y}(t)-{\cal R}_{x,y}(\infty)=\sum_{\lambda}r_{y,x}^{\lambda} e^{-\lambda q t}$. Thus the  equation (\ref{eq.0.12}), for $1\leq x<y<x+L$ gives rise to the following difference equation: $r_{y-1,x}^{\lambda}+r_{y+1,x}^{\lambda}+ r_{y,x-1}^{\lambda}+r_{y,x+1}^{\lambda}=2\left(\frac{\gamma+(y-x)\delta}{q} - \lambda\right)r_{y,x}^{\lambda}$. 
With the notation $E\equiv\frac{q\lambda-\gamma}{\delta}$, this equation admits $r_{y,x}^{\lambda}={\cal A}J_{y-x-E}(2q/\delta)+{\cal B} Y_{y-x-E}(2q/\delta) $ as solution, where ${\cal A}$,  ${\cal B}$ and the spectrum $\{E\}$ are determined from the boundary  and the initial conditions. Indeed, the boundary conditions  ${\cal R}_{x,x}(t)=1$ and $\frac{d}{dt} {\cal R}_{x,x+L}=-L\delta{\cal R}_{x,x+L} $ require respectively, $r_{x,x}^{\lambda}={\cal A}J_{-E}(2q/\delta)+ {\cal B}Y_{-E}(2q/\delta)=0$ and $\sum_{\lambda}\left(q\lambda -\delta L\right)e^{-\lambda q t}\left[{\cal A} J_{L-E}(2q/\delta)+{\cal B} Y_{L-E}(2q/\delta) \right]=0$, i.e.,

\begin{eqnarray}
\label{eq.0.13.1}
\left\{
\begin{array}{l l}
{\cal A}J_{-E}(2q/\delta)+{\cal B}Y_{-E}(2q/\delta)=0\\
{\cal A}J_{L-E}(2q/\delta)+{\cal B}Y_{L-E}(2q/\delta)=0,
\end{array}
\right.
\end{eqnarray}
The only non-trivial solution of this system (for which ${\cal A}\neq 0$ and ${\cal B}\neq 0$) requires
\begin{eqnarray}
\label{eq.0.13.2}
J_{L-E}(2q/\delta)Y_{-E}(2q/\delta)-J_{-E}(2q/\delta)Y_{L-E}(2q/\delta)=0, 
\end{eqnarray}
or equivalently in terms of {\it Lommel function} \cite{tables}:
\begin{eqnarray}
\label{eq.0.13.3}
R_{L-1,1-E}(2i|q|/\delta)=0, 
\end{eqnarray}
Thus the relaxation spectrum of the string-function of the model is obtained as the zeroes of the {\it Lommel function} (\ref{eq.0.13.3}). The latter admits $(L-1)$ zeroes which are symmetrically distributed around $\frac{L}{2}$ (which is also an eigenvalue if $L$ is even) and have a degeneracy $L$. To obtain  the complete set of $L(L-1)+1$ {\it eigenvalues}, i.e., the relaxation spectrum  $\{E_{i}\}, i=1,\dots,L$ of the {\it string-function} (and not the spectrum of the {\it Hamiltonian} (\ref{eq.0.1})), we have also to take into account the {\it eigenvalue} $q\lambda=L\delta$, which follows directly  from the boundary condition $\frac{d}{dt} {\cal R}_{x,x+L}=-L\delta{\cal R}_{x,x+L} $.

To our knowledge there are no explicit results on the zeroes of the Lommel function of imaginary arguments. In order to have more explicit information on the spectrum, we use the formal analogy, first noticed by Peschel et al. \cite{Peschel}, which  exists between the problem under consideration and  the energy spectrum of an electron in a finite one-dimensional crystal in an  electric potential of strength ${\cal E}n$ (here  ${\cal E}=1$)\cite{Stey}.

To compute explicitly $\{E_{i}\},i=1,\dots,L$, we take advantage of the following  eigenvalue-problem:
\begin{eqnarray}
\label{eq.0.19}
\left\{
\begin{array}{l l }
(E-n)F_{n}=V\left(F_{n-1}+F_{n+1}\right)\;;(1\leq n<L)\\
F_{0}=F_{L}=0,
\end{array}
\right.
\end{eqnarray}
where  $F_{n}=V^{E}\left[J_{E-n}(2V)J_{-E}(2V) - (-1)^{n} J_{E}(2V)J_{n-E}(2V)\right]$ are  eigenfunctions. The eigenvalues of (\ref{eq.0.19}) are obtained as the zeroes of the following {\it Lommel function}: $R_{L-1,1-E}(2V)=0$ \cite{Stey}.
Choosing $V=\frac{i|q|}{\delta}$, the problem of determining the relaxation spectrum is reformulated as that of solving the eigenvalue-problem (\ref{eq.0.19}). The latter can be recasted into the following form:
${\cal M}|{\cal F}\rangle\rangle=E|{\cal F}\rangle\rangle$, where ${\cal M}$ is a $(L-1)\times (L-1)$ symmetric (in fact {\it antihermitian}) tridiagonal matrix and $|{\cal F}\rangle\rangle$ is a $(L-1)$-components  column-vector: $|{\cal F}\rangle\rangle\equiv (F_{n=1}\; F_{2}\; \dots  F_{L-1})^{T}$. The general form of the matrix ${\cal M}$ is the following:
\begin{eqnarray}
\label{eq.0.22}
{\cal M} =
\left(
 \begin{array}{c c c c c c c}
 1 &  V &   0  & \dots &\dots & \dots &   0  \\
 V  &  2 &  V &  0 &\dots & \dots &  0  \\
 0 &  V  &  3 &  V & 0 &\dots  &  0  \\
 0 & \ddots &\ddots & \ddots & \ddots  & \ddots & \vdots\\
\vdots &\ddots & \ddots & \ddots & \ddots & \ddots &  0\\
 0 & \dots &  0 & \ddots &   V&  (L-2) &   V\\
 0 & \dots &\dots & \dots & 0 &  V&  (L-1) 
 \end{array}
\right)\
\end{eqnarray}
For {\it small} systems the $(L-1)$ distinct eigenvalues $\{E_{i}\}$ of (\ref{eq.0.22}) can be computed analytically. For $L=6$, we have $\{E_{i}\}=\left\{3,3\pm\sqrt{\frac{5+4V^{2}\pm\sqrt{9+24V^{2}+4V^{4}}}{2}}\right\}$, where we still have to take into account the additional {\it eigenvalue} $q\lambda =L\delta$.
 For larger matrices we had to proceed numerically. Our analysis (based on the spectrum of large  matrices, with $L\leq 1000$), shows that the spectrum $\{E_{i}\}$ (and therefore $\{q\lambda\}$) is {\it real} and symmetric  around $\frac{L}{2}$ which is an eigenvalue when $L$ is even. The other {\it eigenvalues} are not  generally {\it integers}, but for the {\it central} part of the spectrum (when eigenvalues which are close of  $\frac{L}{2}$), the eigenvalues approach integer values. This is not the case at the extremities of the spectrum. In particular, the smallest eigenvalue $E^{\ast}=min_{E_{i}}\{E_{i}\}$ is not an integer and depends on the size of the system:
$E^{\ast}=\epsilon_{L}>1$. However, for $L\gg 1$, $\epsilon_{L}\rightarrow \epsilon_{\infty}$, and  $E^{\ast}$ is a constant: $E^{\ast}=\epsilon_{\infty}>1$. For $L=6$, we have the exact result $\epsilon_{L=6}=3-\sqrt{\frac{5+4V^{2}+\sqrt{9+24V^{2}+4V^{4}}}{2}}$, with $1<\epsilon_{L=6}<3-\frac{1}{2}\sqrt{8+\sqrt{13}}$. This expression can be considered as an excellent approximation to systems of size  $L\gg 1$ and in particular for $\epsilon_{\infty}$. As an illustration, for the case   $\Gamma_{0 0}^{1 0}=3/10$, $\Gamma_{1 0}^{1 1}=1/2$, $\Gamma_{1 1}^{1 0}=1$ and  $\Gamma_{1 0}^{0 0}=1/3$, with the expression above, we obtain (analytically) $\epsilon_{L=6}=1.0823337683$. For larger systems ($L= 10, 25, 40, 1000$), we obtain numerically (with an accuracy of $10^{-10}$): $\epsilon_{10}=\epsilon_{25}=\epsilon_{40}=\epsilon_{1000}=1.0823337697$.
 
 Therefore, the long-time dynamics (of large systems, with $L\gg 1$) is governed by the eigenvalue $E^{\ast}=\epsilon_{L}\approx3-\sqrt{\frac{5+4V^{2}+\sqrt{9+24V^{2}+4V^{4}}}{2}} $ , i.e.
\begin{eqnarray}
\label{eq.0.23}
q\lambda^{\ast}=E^{\ast}\delta+\gamma =\epsilon_{L}\delta+\gamma
=2\left[\frac{\Gamma_{0 0}^{1 0}\Gamma_{1 0}^{1 1}+\Gamma_{1 1}^{1 0}(2\Gamma_{0 0}^{1 0}-
\Gamma_{1 0}^{1 1})}{\Gamma_{0 0}^{1 0}} + (\epsilon_{L}-1)(\Gamma_{0 0}^{1 0}+\Gamma_{1 1}^{1 0})\right]>2\Gamma_{1 0}^{1 1}\geq 0
\end{eqnarray}

The equation (\ref{eq.0.23}) provides the {\it inverse of the relaxation-time} of the system \cite{Mobar}.

With the knowledge of the spectrum $\{E_{i}\}, i=1,\dots, L$, the expression of the density at site $x$ reads:
\begin{eqnarray}
\label{eq.0.14}
\langle n_{x}(t)\rangle - \langle n_{x}(\infty)\rangle=
\frac{\mu}{b}\sum_{E_{i}}{\cal A}_{E_{i}}e^{-(E_{i}\delta+\gamma)t}
\left[Y_{1-E_{i}}(2q/\delta)J_{L-E_{i}}(2q/\delta) -
J_{1-E_{i}}(2q/\delta)Y_{L-E_{i}}(2q/\delta) \right],
\end{eqnarray}
The coefficients ${\cal A}_{E_{i}}$ are obtained from the initial condition (in the translationally-invariant situation, where $S_{x,y}(t)=
S_{y-x}(t)$) according to:
\begin{eqnarray}
\label{eq.0.15}
{\cal A}_{E_{i}}=\sum_{j,n=1}^{L}\left[{\cal N}^{-1}\right]_{i,j}
\left(J_{n-E_{j}}(2q/\delta)Y_{L-E_{j}}(2q/\delta) -Y_{n-E_{j}}(2q/\delta)J_{L-E_{j}}(2q/\delta) \right)^{\ast}\left(S_{n}(0)-S_{n}(\infty)\right)\mu^{-n},
\end{eqnarray}
where ${\cal N}$ is an hermitian $L\times L$ matrix which entries read (for details, see \cite{Mobar}):
${\cal N}_{i,j}=\\
\sum_{n=1}^{L}\left(J_{n-E_{i}}(2q/\delta)Y_{L-E_{i}}(2q/\delta) -Y_{n-E_{i}}(2q/\delta) J_{L-E_{i}}(2q/\delta) \right)^{\ast}\left(J_{n-E_{j}}(2q/\delta)Y_{L-E_{j}}(2q/\delta) -Y_{n-E_{j}}(2q/\delta)J_{L-E_{j}}(2q/\delta) \right) $.

The long-time behaviour of $\langle n_{x}(t)\rangle$ is obtained in retaining in (\ref{eq.0.14}) only the term $E_{i}=E^{\ast}$. 

The result (\ref{eq.0.14}) can be extended to the computation of $\langle n_{x}(t)n_{x_{0}}(0)\rangle$. To do this it suffices to take  $n_{x_{0}}|P(0)\rangle$ (instead of $|P(0)\rangle$) as initial state in (\ref{eq.0.14}) and thus to replace the coefficients ${\cal A}_{E_{i}}$ (\ref{eq.0.15}) by those computed in considering  $\left\langle \{\prod_{j=x}^{y-1}(a-bn_{j}(0))\}  n_{x_{0}}(0)\right\rangle$ instead of $S_{y-x}(0)$.

In this work we have proposed a natural generalization of the IPDF method, to solve (with some restrictions on the reaction-rates) an one-dimensional reaction-diffusion model which can be viewed as an epidemic model and/or a generalization of the voter model and that could not be solved by previous approaches.

On a finite and periodic lattice, we have obtained the exact expression of the steady-state, of the dynamical part of the density and of the non-instantaneous two-point correlation functions of the model under consideration, which exhibits a {\it massive} and {\it real} relaxation spectrum. This means that the steady-state (\ref{eq.0.13}) of the system is reached exponentially with a relaxation-time which is determined explicitly in (\ref{eq.0.23}).

We thank Laurent Klinger for his computational assistance. The support of the Swiss National Fonds is gratefully acknowledged.

%
%
%
%

%

%
\end{document}